\documentclass[12pt,a4paper]{JHEP3}

\title{On the fermionic T-duality of the $AdS_4 \times
  \mathbb{C}\mathrm{P}^3$ sigma-model}

\author{Ido~Adam \\
  Max-Planck-Institut f\"ur Gravitationsphysik (Albert-Einstein-Institut)\\
  Am M\"uhlenberg 1, D-14476 Golm, Germany\\
  \email{idoadam@aei.mpg.de}}

\author{Amit~Dekel and Yaron~Oz \\
  Raymond and Beverly Sackler School of Physics and Astronomy \\
  Tel-Aviv University, Ramat-Aviv 69978, Israel\\
\email{amitde@post.tau.ac.il}, \email{yaronoz@post.tau.ac.il}}

\abstract{In this note we consider a fermionic T-duality of the coset
  realization of the type IIA sigma-model on $AdS_4 \times
  \mathbb{C}\mathrm{P}^3$ with respect to the three flat directions in
  $AdS_4$, six of the fermionic coordinates and three of the
  $\mathbb{C}\mathrm{P}^3$ directions.  We show that the Buscher
  procedure fails as it leads to a singular transformation and discuss
  the result and its implications.
}


\preprint{AEI-2010-128}

\begin{document}

\section{Introduction and summary}

Since the $\mathcal{N} = 6$ superconformal Chern-Simons theory with
matter was proposed by ABJM \cite{Aharony:2008ug} as a dual to
M-theory on $AdS_4 \times S^7/\mathbb{Z}_k$, which reduces in a
certain limit to the type IIA superstring on $AdS_4 \times
\mathbb{C}\mathrm{P}^3$, much work has been devoted to understanding
the properties of the ABJM field theory.

Several tree-level scattering amplitudes of the ABJM theory were
computed \cite{Agarwal:2008pu} and were shown to possess a Yangian
symmetry, which includes the non-local charges and the dual
superconformal symmetry \cite{Bargheer:2010hn}. Some light-like
polygonal Wilson loops in the ABJM theory were computed in
\cite{Henn:2010ps} and hinted that the ABJM theory may have a
scattering amplitudes/Wilson loop duality, which would further support
the case in favor of the existence of dual superconformal
symmetry. Additionally, a contour integral reproducing the known
tree-level amplitudes has been recently proposed and was shown to have
a Yangian symmetry \cite{Lee:2010du}. Furthermore, a differential
representation of a dual superconformal symmetry at tree-level has
been constructed \cite{Huang:2010qy}. This representation involves
variables dual to the ones parameterizing part of the R-symmetry in
addition to the ones dual to the bosonic and fermionic momenta.

The corresponding findings in $\mathcal{N} = 4$ SYM in four dimensions
were explained from the point of view of string theory on $AdS_5
\times S^5$ by a combination of bosonic and fermionic T-dualities,
which is exact at the string tree-level
\cite{Berkovits:2008ic,Beisert:2008iq} (see \cite{Beisert:2009cs} for
a short review). Hence, it is interesting to see whether that is also
the case for type IIA strings on $AdS_4 \times
\mathbb{C}\mathrm{P}^3$.  Previously, it was found that the
sigma-model for $AdS_4 \times \mathbb{C}\mathrm{P}^3$, realized as the
coset $\mathrm{OSp}(6|4)/(\mathrm{SO}(2,1) \times \mathrm{U}(3))$
constructed in \cite{Stefanski:2008ik,Arutyunov:2008if}, was not
self-dual under T-duality involving both three directions in $AdS_4$
and six fermionic coordinates \cite{Adam:2009kt,Grassi:2009yj}. In
fact, one could not perform a fermionic T-duality in six fermionic
isometries which together with the dualized bosonic ones form an
Abelian subgroup of the whole isometry group.

In this note, in light of a suggestion that T-dualizing three
isometries of $\mathbb{C}\mathrm{P}^3$ is also required
\cite{Bargheer:2010hn} and the new evidence
\cite{Lee:2010du,Huang:2010qy} from the field theory, we consider the
fermionic T-duality along the three flat $AdS_4$ coordinates, three
complex Killing vectors in $\mathbb{C}\mathrm{P}^3$ (each one of real
dimension one) as well as six of the fermionic coordinates, whose
corresponding tangent-space vectors generate an Abelian subgroup of
the isometry group. We show that as in the case of dualizing just in
$AdS_4$ and the fermions, the Buscher procedure fails as it leads to a
singular transformation \cite{Adam:2009kt}.

The outline of this note is as follows: in Section
\ref{sec:T-dualizing-AdS4CP3} we apply the Buscher procedure for
T-duality to the $\mathrm{OSp}(6|4)/(\mathrm{SO}(2,1) \times
\mathrm{U}(3))$ Green-Schwarz sigma-model describing type IIA strings
on $AdS_4 \times \mathbb{C}\mathrm{P}^3$ in a certain partial
gauge-fixing and show that it fails. In Section \ref{sec:discussion}
we discuss the implications of the
result. The $\mathrm{osp}(6|4)$ algebra is given in Appendix
\ref{sec:osp64-algebra}.

\section{T-dualizing $AdS_4 \times
  \mathbb{C}\mathrm{P}^3$} \label{sec:T-dualizing-AdS4CP3}

We attempt to T-dualize $AdS_4 \times \mathbb{C}\mathrm{P}^3$ along
the directions corresponding to $P_a$, $Q_{l\alpha}$, $R_{kl}$, which
form an Abelian subalgebra of the isometry group.

We assume that $\kappa$-symmetry can be partially gauge-fixed to set
the six coordinates corresponding to $\hat S^l_\alpha$ to zero and
choose the coset representative
\begin{equation}
  g = e^{x^a P_a + \theta^{l\alpha} Q_{l\alpha}+ y^{kl} R_{kl}} e^B \ , \quad
  e^B = e^{\hat \theta^\alpha_l \hat Q^l_\alpha + \xi^{l\alpha} S_{l\alpha}} y^D
  e^{\hat y_{kl} \hat R^{kl}} \ ,
\end{equation}
where the indices $a = 0, 1, 2$ run over the flat directions of
$AdS_4$, $\alpha = 1, 2$ are $AdS_4$ spinor indices and $l = 1, 2, 3$
are $\mathrm{U}(3)$ fundamental representation indices (see Appendix
\ref{sec:osp64-algebra} for further details). The Maurer-Cartan
one-form is
\begin{equation}
  K = J + j \ , \quad
  J = e^{-B} (dx^a P_a + d\theta^{l\alpha} Q_{l\alpha} + dy^{kl} R_{kl})
  e^B \ , \quad
  j = e^{-B} de^B \ .
\end{equation}
Examining the algebra, one finds that the current $J$ takes values in
the space spanned by $\{ P_a, Q_{l\alpha}, R_{kl}, \hat Q^l_\alpha,
\lambda_k{}^l, \hat R^{kl} \}$, while $j$ is valued in $\mathrm{span}
\{\hat Q^l_\alpha, S_{l \alpha}, \hat S^l_\alpha, D, M_{ab},
\lambda_k{}^l, \hat R^{kl} \}$.

Denoting the decomposition of $K$ into the $\mathbb{Z}_4$-invariant
subspaces by $K_i \in \mathcal{H}_i$, the Green-Schwarz action takes
the form
\begin{eqnarray}
  S & = & \frac{R^2}{4 \pi \alpha'} \int d^2 z \bigg\{ -\frac{1}{2}
  \eta_{ab} J_{P_a} \bar J_{P_b} - j_D \bar j_D - 2 J_{R_{kl}} (\bar
  J_{\hat R^{kl}} + \bar j_{\hat R^{kl}}) - 2 \bar J_{R_{kl}} (J_{\hat
    R^{kl}} + j_{\hat R^{kl}}) - \nonumber \\
  && {} - \frac{i}{2} C_{\alpha \beta} \left[ J_{Q_{l\alpha}} (\bar
    J_{\hat Q^l_\beta} + \bar j_{\hat Q^l_\beta}) - (J_{\hat
      Q^l_\alpha} + j_{\hat Q^l_\alpha}) \bar J_{Q_{l\beta}} -
    j_{S_{l\alpha}} \bar j_{\hat S^l_\beta} + j_{\hat S^l_\alpha} \bar
    j_{S_{l\beta}} \right] \bigg\} \ .
\end{eqnarray}

We attempt to T-dualize the action by using the Buscher procedure
\cite{Buscher:1987sk,Buscher:1987qj} by introducing the new fields
$A^a$, $A^{l\alpha}$, $A^{kl}$, $\bar A^a$, $\bar A^{l\alpha}$ and
$\bar A^{kl}$ such that the current now reads
\begin{equation}
  J = e^{-B} (A^a P_a + A^{l\alpha} Q_{l\alpha} + A^{kl} R_{kl}) e^B
  \ ,
\end{equation}
while $j$, which does not contain $x^a$, $\theta^{l\alpha}$ and
$y^{kl}$, remains unmodified. In addition, the following Lagrange
multiplier terms are added to the action:
\begin{equation}
  S_\mathrm{L} = \frac{R^2}{4 \pi \alpha'} \int d^2 z \left[ \tilde x_a
    (\bar \partial A^a - \partial \bar A^a) + \tilde \theta_{l\alpha}
    (\bar \partial A^{l\alpha} - \partial\bar A^{l\alpha}) + \tilde
    y_{kl} (\bar \partial A^{kl} - \partial \bar A^{kl}) \right] \ ,
\end{equation}
where $\tilde x_a$, $\tilde \theta_{l\alpha}$ and $\tilde y_{kl}$ are
Lagrange multipliers.

The T-duality is performed by integrating out the gauge fields, whose
equations of motion are
\begin{eqnarray}
  0 & = & -\frac{1}{2} \eta_{bc} [e^{-B} P_a e^B]_{P_b} J_{P_c} +
  \frac{i}{2} C_{\alpha \beta} \Big[ [e^{-B} P_a e^B]_{Q_{l\alpha}}
    (J_{\hat Q^l_\beta} + j_{\hat Q^l_\beta}) - \nonumber \\
    && {} - [e^{-B} P_a e^B]_{\hat Q^l_\alpha} J_{Q_{l\beta}} \Big] -
  2 [e^{-B} P_a e^B]_{R_{kl}} (J_{\hat R^{kl}} + j_{\hat R^{kl}}) - 2
  [e^{-B} P_a e^B]_{\hat R^{kl}} J_{R_{kl}} + \partial \tilde x_a \ ,
  \nonumber \\
  0 & = & -\frac{1}{2} \eta_{bc} [e^{-B} Q_{l\alpha} e^B]_{P_b} J_{P_c} +
  \frac{i}{2} C_{\beta \gamma} \Big[ [e^{-B} Q_{l\alpha} e^B]_{Q_{k\beta}}
    (J_{\hat Q^k_\gamma} + j_{\hat Q^k_\gamma}) - \nonumber \\
    && {} - [e^{-B} Q_{l\alpha} e^B]_{\hat Q^k_\beta} J_{Q_{k\gamma}} \Big] -
  2 [e^{-B} Q_{l\alpha} e^B]_{R_{pq}} (J_{\hat R^{pq}} + j_{\hat R^{pq}}) - 2
  [e^{-B} Q_{l\alpha} e^B]_{\hat R^{pq}} J_{R_{pq}} - \nonumber \\
  && {} - \partial \tilde \theta_{l\alpha} \ , \nonumber \\
  0 & = & -\frac{1}{2} \eta_{bc} [e^{-B} R_{kl} e^B]_{P_b} J_{P_c} +
  \frac{i}{2} C_{\alpha \beta} \Big[ [e^{-B} R_{kl} e^B]_{Q_{p\alpha}}
    (J_{\hat Q^p_\beta} + j_{\hat Q^p_\beta}) - \nonumber \\
    && {} - [e^{-B} R_{kl} e^B]_{\hat Q^p_\alpha} J_{Q_{p\beta}} \Big] -
  2 [e^{-B} R_{kl} e^B]_{R_{pq}} (J_{\hat R^{pq}} + j_{\hat R^{pq}}) - 2
  [e^{-B} R_{kl} e^B]_{\hat R^{pq}} J_{R_{pq}} + \nonumber \\
  && {} + \partial \tilde y_{kl}
\end{eqnarray}
for the holomorphic fields and
\begin{eqnarray}
  0 & = & -\frac{1}{2} \eta_{bc} [e^{-B} P_a e^B]_{P_b} \bar J_{P_c} -
  \frac{i}{2} C_{\alpha \beta} \Big[ [e^{-B} P_a e^B]_{Q_{l\alpha}}
    (\bar J_{\hat Q^l_\beta} + \bar j_{\hat Q^l_\beta}) - [e^{-B} P_a
      e^B]_{\hat Q^l_\alpha} \bar J_{Q_{l\beta}} \Big] - \nonumber \\
  && {}  - 2 [e^{-B} P_a e^B]_{R_{kl}} (\bar J_{\hat R^{kl}} + \bar j_{\hat R^{kl}}) - 2
  [e^{-B} P_a e^B]_{\hat R^{kl}} \bar J_{R_{kl}} - \bar \partial \tilde x_a \ ,
  \nonumber \\
  0 & = & -\frac{1}{2} \eta_{bc} [e^{-B} Q_{l\alpha} e^B]_{P_b} \bar J_{P_c} -
  \frac{i}{2} C_{\beta \gamma} \Big[ [e^{-B} Q_{l\alpha} e^B]_{Q_{k\beta}}
    (\bar J_{\hat Q^k_\gamma} + \bar j_{\hat Q^k_\gamma}) - \nonumber \\
    && {} - [e^{-B} Q_{l\alpha} e^B]_{\hat Q^k_\beta} \bar J_{Q_{k\gamma}} \Big] -
  2 [e^{-B} Q_{l\alpha} e^B]_{R_{pq}} (\bar J_{\hat R^{pq}} + \bar j_{\hat R^{pq}}) - 2
  [e^{-B} Q_{l\alpha} e^B]_{\hat R^{pq}} \bar J_{R_{pq}} + \nonumber \\
  && {} + \bar \partial \tilde \theta_{l\alpha} \ , \nonumber \\
  0 & = & -\frac{1}{2} \eta_{bc} [e^{-B} R_{kl} e^B]_{P_b} \bar J_{P_c} -
  \frac{i}{2} C_{\alpha \beta} \Big[ [e^{-B} R_{kl} e^B]_{Q_{p\alpha}}
    (\bar J_{\hat Q^p_\beta} + \bar j_{\hat Q^p_\beta}) - \nonumber \\
    && {} - [e^{-B} R_{kl} e^B]_{\hat Q^p_\alpha} \bar J_{Q_{p\beta}} \Big] -
  2 [e^{-B} R_{kl} e^B]_{R_{pq}} (\bar J_{\hat R^{pq}} + \bar j_{\hat R^{pq}}) - 2
  [e^{-B} R_{kl} e^B]_{\hat R^{pq}} \bar J_{R_{pq}} - \nonumber \\
  && {} - \bar \partial \tilde y_{kl}
\end{eqnarray}
for the anti-holomorphic ones. (The complexity of the equations arises
from the fact that, unlike in the $AdS_5 \times S^5$ case, $J$ is valued
in a space larger than the one that is actually dualized.)

For the purpose of solving these equations, the properties of the
field-dependent group-theoretic factors must be understood. In
particular, it should be checked whether the coefficients of the gauge
fields have non-trivial kernels.

In order to do so, we resort to explicitly expressing the currents in
terms of the coordinates. We denote $C \equiv \hat \theta^\alpha_l \hat
Q^l_\alpha + \xi^{l\alpha} S_{l\alpha}$ and examine the commutators
\begin{eqnarray}
  [P_a, C] & = & -\frac{i}{\sqrt{2}} \gamma_{a\alpha}{}^\beta \xi^{l\alpha}
  Q_{l\beta} \equiv \Xi^{Pl\beta}_a Q_{l\beta} \ , \nonumber \\
  {}[Q_{l\beta}, C] & = & \frac{1}{\sqrt{2}} (\gamma^a C)_{\beta\alpha}
  \hat \theta^\alpha_l P_a + \frac{1}{\sqrt{2}} C_{\beta\alpha}
  \xi^{k\alpha} R_{lk} \equiv \Theta^{Qa}_{l\beta} P_a +
  \Xi^{Qk}_\beta R_{lk} \equiv M_{l\beta} \ , \nonumber \\
  {}[R_{kl}, C] & = & -\frac{i}{\sqrt{2}} (\hat \theta^\alpha_l
  \delta^p_k - \hat \theta^\alpha_k \delta^p_l) Q_{p\alpha} \equiv
  \Theta^{Rp\alpha}_{kl} Q_{p\alpha} \ .
\end{eqnarray}
We further define
\begin{equation}
  N_{l\alpha}{}^{k\beta} = \Theta^{Qa}_{l\alpha} \Xi^{P k\beta}_a +
  \Xi^{Qp}_\alpha \Theta^{Rk\beta}_{pl}
\end{equation}
and note that $[M_{l\alpha}, C] = N_{l\alpha}{}^{k\beta} Q_{k\beta}$
and $[Q_{l\alpha}, C] = M_{l\alpha}$. Using the formula $e^{-B} A e^B
= A + [A, B] + \frac{1}{2!} [[A, B], B] + \dots$, we get
\begin{eqnarray}
  \lefteqn{e^{-C} (dx^a P_a + d\theta^{l\alpha} Q_{l\alpha} + dy^{kl}
    R_{kl}) e^C = dx^a P_a  + dy^{kl} R_{kl} +}
  \nonumber \\
  && {} + (dx^a \Xi^{Pl\alpha}_a + dy^{pq} \Theta^{Rl\alpha}_{pq})
  \left[ {\left( \frac{\cosh\sqrt{N} - 1}{N}
      \right)_{l\alpha}}^{k\beta} M_{k\beta} + {\left(
      \frac{\sinh\sqrt{N}}{\sqrt{N}} \right)_{l\alpha}}^{k\beta}
    Q_{k\beta}  \right] +  \nonumber \\
  && {} + d\theta^{l\alpha} \left[ {\left( \frac{\sinh
        \sqrt{N}}{\sqrt{N}}\right)_{l\alpha}}^{k\beta} M_{k\beta} +
    {(\cosh\sqrt{N})_{l\alpha}}^{k\beta} Q_{k\beta} \right] \ .
\end{eqnarray}
Finally, conjugating with $y^D e^{\hat y_{kl} \hat R^{kl}}$ yields
the current
\begin{eqnarray}
  \lefteqn{J = \frac{dx^a}{y} P_a + dy^{kl} (R_{kl} + 2 i \sqrt{2}
    \hat y_{kq} {\lambda_l}^q + 2 \hat y_{kq} \hat y_{ln} \hat R^{qn})
  + } \nonumber \\
  && + \left[ (dx^a \Xi^{Pl\alpha}_a + dy^{pq} \Theta^{Rl\alpha}_{pq})
    {\left( \frac{\cosh\sqrt{N} - 1}{N} \right)_{l\alpha}}^{k\beta} +
    d\theta^{l\alpha} {\left(\frac{\sinh\sqrt{N}}{\sqrt{N}}
      \right)_{l\alpha}}^{k\beta} \right] \times \nonumber \\
  && {} \times \left[ \tilde M_{k\beta} + i \sqrt{2} \Xi^{Qm}_\beta
    (\hat y_{kq} {\lambda_m}^q - \hat y_{mq} {\lambda_k}^q) +
    \Xi^{Qr}_\beta (\hat y_{kq} \hat y_{rn} - \hat y_{rq}
    \hat y_{kn}) \hat R^{qn} \right] + \nonumber \\
  && {} + \frac{1}{y^{1/2}} \left[ (dx^a \Xi^{Pl\alpha}_a + dy^{pq}
    \Theta^{Rl\alpha}_{pq}) {\left( \frac{\sinh\sqrt{N}}{\sqrt{N}}
      \right)_{l\alpha}}^{k\beta} + d\theta^{l\alpha}
          {(\cosh\sqrt{N})_{l\alpha}}^{k\beta} \right] \times
  \nonumber \\
  && \times (Q_{k\beta} + i \sqrt{2} \hat y_{pk} \hat Q^p_\beta) \ ,
\end{eqnarray}
where $\tilde M_{k\beta} \equiv y^{-D} M_{k\beta} y^D = \frac{1}{y}
\Theta^{Qa}_{l\alpha} P_a + \Xi^{Ql}_\alpha R_{kl}$.

Unfortunately, $j$ is even more complicated. However, before plunging
into its computation in a closed form it is worthwhile to examine it
to the lowest order in $\hat \theta^\alpha_l$ and
$\xi^{l\alpha}$. Doing so yields,
\begin{equation}
  j = \frac{d\hat \theta^\alpha_l}{y^{1/2}} \hat Q^l_\alpha + y^{1/2}
  d\xi^{l\alpha} S_{l\alpha} - i \sqrt{2} y^{1/2} \hat y_{kl}
  d\xi^{l\alpha} \hat S^k_\alpha + \frac{dy}{y} D + d\hat y_{pq} \hat
  R^{pq} + O(\hat \theta^\alpha_l, \xi^{l\alpha}) \ .
\end{equation}

Having the currents, we can take a look at the action to lowest order
in $\hat \theta^\alpha_l$ and $\xi^{l\alpha}$:
\begin{eqnarray}
  S & = & \frac{R^2}{4 \pi \alpha'} \int d^2 z \bigg\{ -\frac{1}{2}
  \eta_{ab} \frac{\partial x^a \bar \partial x^b}{y^2} -
  \frac{\partial y \bar \partial y}{y^2} - 2 \partial y^{kl} (2 \hat
  y_{pk} \hat y_{ql} \bar \partial y^{pq} + \bar \partial \hat y_{kl})
  - \nonumber \\
  && {} - 2 \bar \partial y^{kl} (2 \hat y_{pk} \hat y_{ql} \partial
  y^{pq} + \partial \hat y_{kl}) - \frac{i}{2 y} C_{\alpha \beta}
  \Big[ \partial \theta^{l\alpha} (i \sqrt{2} \hat y_{kl} \bar
    \partial \theta^{k\beta} + \bar \partial \hat \theta^\beta_l) -
    \nonumber \\
    && {} - (i \sqrt{2} \hat y_{kl} \partial \theta^{k\alpha} +
    \partial \hat \theta^\alpha_l) \bar \partial \theta^{l \beta}
    \Big] + \frac{i}{2} y C_{\alpha \beta} (-i \sqrt{2} \hat y_{lk}
  \partial \xi^{l\alpha} \bar \partial \xi^{k\beta} + i \sqrt{2} \hat
  y_{lk} \partial \xi^{k\alpha} \bar \partial \xi^{l\beta}) \bigg\}
  \ . \nonumber \\
\end{eqnarray}
The term quadratic in the $\theta^{l\alpha}$ derivatives is multiplied
by a three-dimensional antisymmetric matrix, whose rank is two, and
the higher order terms in $\hat \theta^\alpha_l$ and $\xi^{l\alpha}$
cannot make the matrix's kernel trivial.  Thus the term quadratic in the
fermionic gauge fields in the dualized action will be multiplied by a
singular matrix and the fermionic gauge fields will be multiplied by a
singular matrix in the equations of motion --- one cannot T-dualize
all the six fermionic coordinates.

Since the obstruction to T-dualizing the fermionic coordinates is at
the zeroth order in the spectator fermions, it appears that modifying
the $\kappa$-symmetry gauge-fixing of these fermionic degrees of
freedom would not change the above conclusion.

\section{Discussion} \label{sec:discussion}

We showed that the application of the Buscher T-duality procedure to
the coset $\mathrm{OSp}(6|4)/(\mathrm{SO}(2,1) \times \mathrm{U}(3))$
fails when dualizing along the $AdS_4$ flat directions, three of the
(real) $\mathbb{C}\mathrm{P}^3$ directions and six fermionic
directions.  There are several ways to explain this apparent tension
between the field theory tree-level evidence and the sigma-model
analysis.

The simplest and most obvious explanation is that the dual
superconformal symmetry exists only in the weakly-coupled field theory
description and breaks down at the strong-coupling regime, which is
described by the string theory dual. A second possibility is that in
this case the dual superconformal symmetry is not related to the
ordinary superconformal symmetry by a T-duality transformation but in
a more intricate way.

A third possibility is that the coset formulation does not capture
the entire superstring description. The coset is obtained by a partial
gauge-fixing of the $\kappa$-symmetry of the full $AdS_4 \times
\mathbb{C}\mathrm{P}^3$ sigma-model \cite{Gomis:2008jt} by setting the
fermionic coordinates corresponding to the eight broken supersymmetries
to zero. However, as noted in \cite{Gomis:2008jt}, this gauge-fixing
is not compatible with all the possible string configurations. Thus, it
does not have a representation for certain field theory operators,
which might amount to a (possibly inconsistent) truncation of the field
theory that does not preserve the dual superconformal symmetry.
A way to resolve this issue could be to use a better gauge-fixing
of the $\kappa$-symmetry as proposed in \cite{Gomis:2008jt,Grassi:2009yj}.

\acknowledgments We would like to thank Y-t.~Huang and A.~E.~Lipstein
for sharing a draft of their paper \cite{Huang:2010qy} with us before its
publication. I.A.\ is supported in part by the German-Israeli Project
cooperation (DIP H.52) and the German-Israeli Fund (GIF).

\appendix

\section{The $\mathrm{osp}(6|4)$ superalgebra} \label{sec:osp64-algebra}

The $\mathrm{osp}(6|4)$ algebra's commutation relations in the so$(1,2)\oplus$u$(3)$ basis are given by
\begin{equation}
[\lambda_{k}{}^{  l},\lambda_{m}{}^{  n}]=\frac{i}{\sqrt{2}}(\delta_{m}{}^{  l}\lambda_{k}{}^{  n}-\delta_{k}{}^{  n}\lambda_{m}{}^{  l})
,\quad
\end{equation}
\begin{equation}
[\lambda_{k}{}^{  l},R_{m n}]=\frac{i}{\sqrt{2}}(\delta_{m}{}^{  l}R_{k n}-\delta_{n}{}^{  l}R_{k m}),\quad
[\lambda_{l}{}^{  k},\hat R^{p q}]=-\frac{i}{\sqrt{2}}(\delta^{p}_{  l}\hat R^{k q}-\delta^{q}_{  l}\hat R^{k p})
\end{equation}
\begin{equation}
[R_{mn},R_{kl}]=0,\qquad
[R_{mn},\hat R^{  k  l}]=\frac{i}{\sqrt{2}}
(\delta_{m}{}^{  k}\lambda_{n}{}^{  l}
-\delta_{m}{}^{  l}\lambda_{n}{}^{  k}
-\delta_{n}{}^{  k}\lambda_{m}{}^{  l}
+\delta_{n}{}^{  l}\lambda_{m}{}^{  k})
\end{equation}
\begin{equation}
[P_a,P_b]=0,\qquad [K_a,K_b]=0, \qquad
[P_a,K_b]=\eta_{ab}D-M_{ab}
\end{equation}
\begin{equation}
[M_{ab},M_{cd}]=
 \eta_{ac}M_{bd}
+\eta_{bd}M_{ac}
-\eta_{ad}M_{bc}
-\eta_{bc}M_{ad}
\end{equation}
\begin{equation}
[M_{ab},P_c]=\eta_{ac}P_b-\eta_{bc}P_a, \qquad
[M_{ab},K_c]=\eta_{ac}K_b-\eta_{bc}K_a
\end{equation}
\begin{equation}
[D,P_a]=P_a, \qquad
[D,K_a]=-K_a, \qquad
[D,M_{ab}]=0
\end{equation}
\begin{equation}
[D,Q_{l\alpha}]=\frac{1}{2} Q_{l\alpha}, \qquad [D,S_{l\alpha}]=-\frac{1}{2} S_{l\alpha}
\end{equation}
\begin{equation}
[P_a,Q_{l\alpha}]=0, \qquad [K_a,S_{l\alpha}]=0
\end{equation}
\begin{equation}
[P_a,S_{l\alpha}]=-\frac{i}{\sqrt{2}}(\gamma_a)_{\alpha}{}^{\beta}Q_{l\beta}, \qquad [K_a,Q_{l\alpha}]=\frac{i}{\sqrt{2}}(\gamma_a)_{\alpha}{}^{\beta}S_{l\beta}
\end{equation}
\begin{equation}
[M_{ab},Q_{l\alpha}]=-\frac{i}{2}(\gamma_{ab})_{\alpha}{}^{\beta}Q_{l\beta}, \qquad
[M_{ab},S_{l\alpha}]=-\frac{i}{2}(\gamma_{ab})_{\alpha}{}^{\beta}S_{l\beta}
\end{equation}
\begin{equation}
[R_{kl},\hat Q^{  p}_\alpha]=\frac{i}{\sqrt{2}}(\delta_l{}^{  p}Q_{k\alpha}-\delta_k{}^{  p}Q_{l\alpha})
,\qquad
[R_{kl},\hat S^{  p}_\alpha]=-\frac{i}{\sqrt{2}}(\delta_l{}^{  p}S_{k\alpha}-\delta_k{}^{  p}S_{l\alpha})
\end{equation}
\begin{equation}
[\hat R^{  k  l},Q_{p\alpha}]=-\frac{i}{\sqrt{2}}(\delta_{p}{}^{   l}\hat Q^{  k}_\alpha-\delta_{p}{}^{   k}\hat Q^{  l}_\alpha)
,\qquad
[\hat R^{  k  l},S_{ p\alpha}]=\frac{i}{\sqrt{2}}(\delta_{ p}{}^{   l}\hat S^{  k}_\alpha-\delta_{p}{}^{   k}\hat S^{  l}_\alpha)
\end{equation}
\begin{equation}
[\lambda_{k}{}^{  l},Q_{ p\alpha}]=\frac{i}{\sqrt{2}}\delta_{p}{}^{   l}Q_{k\alpha}, \qquad
[\lambda_{k}{}^{  l},S_{ p\alpha}]=\frac{i}{\sqrt{2}}\delta^{p}{}^{   l}S_{k\alpha}
\end{equation}
\begin{equation}
[\lambda_{k}{}^{  l},\hat Q^{  p}_\alpha]=-\frac{i}{\sqrt{2}}\delta_k{}^{  p }\hat Q^{  l}_\alpha, \qquad
[\lambda_{k}{}^{  l},\hat S^{  p}_\alpha]=-\frac{i}{\sqrt{2}}\delta_k{}^{  p }\hat S^{  l}_\alpha
\end{equation}
\begin{equation}
\{Q_{l\alpha},Q_{k\beta}\}=0, \qquad
\{Q_{l\alpha},\hat Q^{  k}_\beta\}=-\frac{1}{\sqrt{2}}\delta_{l}{}^{  k}(\gamma^a C)_{\alpha\beta}P_a
\end{equation}
\begin{equation}
\{S_{l\alpha},S_{k\beta}\}=0, \qquad
\{S_{l\alpha},\hat S^{  k}_\beta\}=-\frac{1}{\sqrt{2}}\delta_{l}{}^{  k}(\gamma^a C)_{\alpha\beta}K_a
\end{equation}
\begin{equation}
\{Q_{l\alpha},S_{k\beta}\}=-\frac{1}{\sqrt{2}}C_{\alpha\beta}R_{lk}, \qquad
\{\hat Q^{  l}_\alpha,\hat S^{  k}_\beta\}=-\frac{1}{\sqrt{2}}C_{\alpha\beta}\hat R^{  l  k}
\end{equation}
\begin{equation}
\{Q_{l\alpha},\hat S^{  k}_\beta\}=-i\frac{1}{2}\delta_{l}{}^{  k}(C_{\alpha\beta}D+i\frac{1}{2}(\gamma^{ab}C)_{\alpha\beta}M_{ab})+\frac{1}{\sqrt{2}}C_{\alpha\beta}\lambda_{l}{}^{  k}
\end{equation}
\begin{equation}
\{\hat Q^{  l}_\alpha,S_{k\beta}\}=i\frac{1}{2}\delta_k{}^{  l}(C_{\alpha\beta}D-i\frac{1}{2}(\gamma^{ab}C)_{\alpha\beta}M_{ab})+\frac{1}{\sqrt{2}}C_{\alpha\beta}\lambda_{ k}{}^{   l}
\end{equation}
The indices take the values $k,l=1,...,3$, the $\mathbf{3}$ $\mathrm{u}(3)$, $a,b=0,1,2$ are the $\mathbf{3}$ of $\mathrm{so}(1,2)$ and $\alpha,\beta,..=1,2$ are the $\mathrm{so}(2,1)$ spinors, and $\eta=\textrm{diag}(-,+,+)$. The generators satisfy the following relations under complex conjugation $R^*_{kl}=\hat R^{  k  l}$, $\lambda_{k}{}^{  l}=\lambda^*_{l}{}^{  k}$, $\hat Q^{  l}_\alpha=(Q_{l\alpha})^*$ and $\hat S^l_{\alpha}=(S_{l\alpha})^*$.
The $(\gamma_a)_\alpha{}^\beta$ are the Dirac matrices of $\mathrm{so}(1,2)$, and $\gamma_{ab}=\frac{i}{2}[\gamma_a,\gamma_b]$.
We raise and lower spinor indices using $C_{\alpha\beta}=\epsilon_{\alpha\beta}$, $\psi_\alpha=\psi^\beta\epsilon_{\beta\alpha}$, $\psi^\alpha=\epsilon^{\alpha\beta}\psi_\beta$, where $\epsilon_{12}=-\epsilon_{21}=\epsilon^{12}=-\epsilon^{21}=1$.\\
The bilinear forms are given by
\begin{equation}\label{bilinear_forms_osp64}
\begin{array}{l}
\mathrm{Str} (R_{kl}, \hat R^{  p  q})=\delta_{k}{}^{  q}\delta_{l}{}^{  p}-\delta_{k}{}^{  p}\delta_{l}{}^{  q},\\
\mathrm{Str} (\lambda_{k}{}^{  l} ,\lambda_{p}{}^{  q})=-\delta_{k}{}^{  q}\delta_{l}{}^{  p},\\
\mathrm{Str} (Q_{l\alpha} ,\hat S^{  k}_{\beta})=i\delta_{l}{}^{  k}C_{\alpha\beta},\\
\mathrm{Str} (S_{l\alpha} ,\hat Q^{  k}_{\beta})=-i\delta_k{}^{  l}C_{\alpha\beta},\\
\mathrm{Str} (P_a ,K_b)=-\eta_{ab},\\
\mathrm{Str} (D, D)=-1,\\
\mathrm{Str} (M_{ab}, M_{cd})=\eta_{ac}\eta_{bd}-\eta_{ad}\eta_{bc}.
\end{array}
\end{equation}

The $\mathbb{Z}_4$ subspaces with the invariant locus of
$\mathrm{U}(3)\times \mathrm{SO}(3,1)$ which gives the
semi-symmetric space $AdS_4\times \mathbb{C}\mathrm{P}^3$ are
\begin{equation}
\begin{array}{l}
\mathcal{H}_0=\{P_a-K_a,M_{ab},\lambda_k{}^l\},\\
\mathcal{H}_1=\{Q_{l\alpha}-S_{l\alpha},\hat Q^{  l}_{\alpha}-\hat S^{  l}_{\alpha}\},\\
\mathcal{H}_2=\{P_a+K_a,D,R_{kl},\hat R^{kl}\},\\
\mathcal{H}_3=\{Q_{l\alpha}+S_{l\alpha},\hat Q^{  l}_{\alpha}+\hat S^{  l}_{\alpha}\}.
\end{array}
\end{equation}

\bibliographystyle{utphys}
\bibliography{ido-references}

\providecommand{\href}[2]{#2}\begingroup\raggedright\begin{thebibliography}{10}

\bibitem{Aharony:2008ug}
O.~Aharony, O.~Bergman, D.~L. Jafferis, and J.~Maldacena, ``{N=6 superconformal
  Chern-Simons-matter theories, M2-branes and their gravity duals},''
  \href{http://dx.doi.org/10.1088/1126-6708/2008/10/091}{{\em JHEP} {\bfseries
  10} (2008) 091},
\href{http://arxiv.org/abs/0806.1218}{{\ttfamily arXiv:0806.1218 [hep-th]}}.

\bibitem{Agarwal:2008pu}
A.~Agarwal, N.~Beisert, and T.~McLoughlin, ``{Scattering in Mass-Deformed $N
  \ge 4$ Chern-Simons Models},''
  \href{http://dx.doi.org/10.1088/1126-6708/2009/06/045}{{\em JHEP} {\bfseries
  06} (2009) 045},
\href{http://arxiv.org/abs/0812.3367}{{\ttfamily arXiv:0812.3367 [hep-th]}}.

\bibitem{Bargheer:2010hn}
T.~Bargheer, F.~Loebbert, and C.~Meneghelli, ``{Symmetries of Tree-level
  Scattering Amplitudes in N=6 Superconformal Chern-Simons Theory},''
\href{http://arxiv.org/abs/1003.6120}{{\ttfamily arXiv:1003.6120 [hep-th]}}.

\bibitem{Henn:2010ps}
J.~M. Henn, J.~Plefka, and K.~Wiegandt, ``{Light-like polygonal Wilson loops in
  3d Chern-Simons and ABJM theory},''
\href{http://arxiv.org/abs/1004.0226}{{\ttfamily arXiv:1004.0226 [hep-th]}}.

\bibitem{Lee:2010du}
S.~Lee, ``{Yangian Invariant Scattering Amplitudes in Super-Chern- Simons
  Theory},''
\href{http://arxiv.org/abs/1007.4772}{{\ttfamily arXiv:1007.4772 [hep-th]}}.

\bibitem{Huang:2010qy}
Y.-t. Huang and A.~E. Lipstein, ``{Dual Superconformal Symmetry of N=6
  Chern-Simons Theory},''
\href{http://arxiv.org/abs/1008.0041}{{\ttfamily arXiv:1008.0041 [hep-th]}}.

\bibitem{Berkovits:2008ic}
N.~Berkovits and J.~Maldacena, ``{Fermionic T-Duality, Dual Superconformal
  Symmetry, and the Amplitude/Wilson Loop Connection},''
  \href{http://dx.doi.org/10.1088/1126-6708/2008/09/062}{{\em JHEP} {\bfseries
  09} (2008) 062},
\href{http://arxiv.org/abs/0807.3196}{{\ttfamily arXiv:0807.3196 [hep-th]}}.

\bibitem{Beisert:2008iq}
N.~Beisert, R.~Ricci, A.~A. Tseytlin, and M.~Wolf, ``{Dual Superconformal
  Symmetry from AdS5 x S5 Superstring Integrability},''
\href{http://arxiv.org/abs/0807.3228}{{\ttfamily arXiv:0807.3228 [hep-th]}}.

\bibitem{Beisert:2009cs}
N.~Beisert, ``{T-Duality, Dual Conformal Symmetry and Integrability for Strings
  on $AdS_5 \times S^5$},''
  \href{http://dx.doi.org/10.1002/prop.200900060}{{\em Fortschr. Phys.}
  {\bfseries 57} (2009) 329--337},
\href{http://arxiv.org/abs/0903.0609}{{\ttfamily arXiv:0903.0609 [hep-th]}}.

\bibitem{Stefanski:2008ik}
j.~Stefanski, B., ``{Green-Schwarz action for Type IIA strings on $AdS_4\times
  CP^3$},'' \href{http://dx.doi.org/10.1016/j.nuclphysb.2008.09.015}{{\em Nucl.
  Phys.} {\bfseries B808} (2009) 80--87},
\href{http://arxiv.org/abs/0806.4948}{{\ttfamily arXiv:0806.4948 [hep-th]}}.

\bibitem{Arutyunov:2008if}
G.~Arutyunov and S.~Frolov, ``{Superstrings on $AdS_4 x CP^3$ as a Coset
  Sigma-model},'' \href{http://dx.doi.org/10.1088/1126-6708/2008/09/129}{{\em
  JHEP} {\bfseries 09} (2008) 129},
\href{http://arxiv.org/abs/0806.4940}{{\ttfamily arXiv:0806.4940 [hep-th]}}.

\bibitem{Adam:2009kt}
I.~Adam, A.~Dekel, and Y.~Oz, ``{On Integrable Backgrounds Self-dual under
  Fermionic T- duality},''
  \href{http://dx.doi.org/10.1088/1126-6708/2009/04/120}{{\em JHEP} {\bfseries
  04} (2009) 120},
\href{http://arxiv.org/abs/0902.3805}{{\ttfamily arXiv:0902.3805 [hep-th]}}.

\bibitem{Grassi:2009yj}
P.~A. Grassi, D.~Sorokin, and L.~Wulff, ``{Simplifying superstring and D-brane
  actions in AdS(4) x CP(3) superbackground},''
  \href{http://dx.doi.org/10.1088/1126-6708/2009/08/060}{{\em JHEP} {\bfseries
  08} (2009) 060},
\href{http://arxiv.org/abs/0903.5407}{{\ttfamily arXiv:0903.5407 [hep-th]}}.

\bibitem{Buscher:1987sk}
T.~H. Buscher, ``{A Symmetry of the String Background Field Equations},''
\href{http://dx.doi.org/10.1016/0370-2693(87)90769-6}{{\em Phys. Lett.}
  {\bfseries B194} (1987) 59}.

\bibitem{Buscher:1987qj}
T.~H. Buscher, ``{Path Integral Derivation of Quantum Duality in Nonlinear
  Sigma Models},''
\href{http://dx.doi.org/10.1016/0370-2693(88)90602-8}{{\em Phys. Lett.}
  {\bfseries B201} (1988) 466}.

\bibitem{Gomis:2008jt}
J.~Gomis, D.~Sorokin, and L.~Wulff, ``{The complete AdS(4) x CP(3) superspace
  for the type IIA superstring and D-branes},''
\href{http://arxiv.org/abs/0811.1566}{{\ttfamily arXiv:0811.1566 [hep-th]}}.

\end{thebibliography}\endgroup

\end{document}